**Clickers in the '60s at a TYC?**

**David Marasco,** Foothill College, Los Altos Hills, CA

marascodavid@foothill.edu

At an AAPT conference years ago, a person noticed my badge, exclaiming "I always wanted to work at Foothill College!" They told me that in the 1960s, Foothill had a lecture hall where students as a class could give immediate feedback to their instructors via buttons at their seats, a forerunner to modern clicker systems. Intrigued, I later did a quick Google search and discovered that Foothill had an EDEX system in a 240-person lecture hall, which featured multiple-choice buttons at each seat that were networked to a computer at the instructor's section.[1] This fact got tucked away, and I didn't think about it again until a recent hiring interview when a candidate from an R1 commented how TYCs were at the forefront of innovations in teaching. I remembered the EDEX system, and it reignited my curiosity.

The fundamentals of the EDEX system were student response stations consisting of four buttons, and meters at the instructor's station that showed the relative response rate (e.g., 20% A, 55% B, 15% C, and 10% D). Associated recording devices could be used to record each individual's response, and in conjunction with a seating chart could be used for both grading and attendance. Standard installations were for 10, 20, 30, and 40 stations.[1,2] Edex Teaching Machines was founded in 1962 by Eugene Kleiner, a prominent figure in Silicon Valley history. He is more famous for being one of the founders of Fairchild Semiconductor and the venture capital firm Kleiner, Perkins, Caufield, and Byers.[3] Foothill benefited from our location, as Edex's facility was roughly 5 miles from our campus.[2] Edex was bought by Raytheon in the mid-1960s.

The reference to a 240-seat lecture hall was a source of confusion; we don't currently have a room that size, and while there have been renovations, we have not replaced any buildings. The October 1, 1965, edition of our school newspaper provided the solution, describing the "'Push-Button' Forum Building" that was nearing completion. In the time between its building and my arrival, the room lost 40 seats to a pair of aisles. In addition to the EDEX system, the high-tech classroom featured closed-circuit television so students could view things broadcast from other places on campus, and an amplified telephone that would allow a professor to interview an outside expert while the class listened in.[4] Foothill's system was further described in a paper written by one of our Fine Arts instructors. The lecture hall itself was used by both the Social Science and the Physical Science divisions. The student response buttons were in the armrests of the chairs, and the system was integrated into an audio system that featured a tape player. In addition to the previously mentioned amplified telephone, the instructor had a dedicated phone to talk to an AV technician who supported the room. It was noted that instructors were still developing their skills with regard to how to best use their new tools.[5]

Further research found that EDEX systems were in use at Meramec Community College, Henry Ford Community College, Monterey Peninsula College, and Los Angeles Trade-Technical College (LATTC), while Mt. San Jacinto College built a similar system in-house. While LATTC eventually purchased two 50-seat installations, they had originally written a $7,400 proposal for a 40-seat system.[6,7] In 1968 Raytheon and the Ford Motor Corporation collaborated on a training program for mechanics using EDEX. It featured slide projectors and a prerecorded audio track. Trainees would watch the lectures, and at key points

would be asked to respond to multiple choice questions. If the class as a whole dropped below a benchmark, a teacher would stop the recording and provide live instruction to reinforce the learning.[8]

Did the EDEX system make an impact at Foothill? Sadly, I think the answer is no. Despite a plea to our retirees association, I could not find a faculty member who used it to teach. While this may speak to the survival rates of teachers who were active in the 1960s, I also suspect that the system was not in use for long. One of the proposed drawbacks of the system was the possibility of breakdown.[1] While looking for reports of the EDEX system in Foothill's campus newspaper, I found an article on our Foucault pendulum.[9] Its remains were sitting in a storeroom; it lived for less than 4 years. It was the world's first outdoor installation and met its end when local children rode it like a swing. It was an interesting experiment, but there was little funding for maintenance, as is often the case at two-year colleges. While it was decades ahead of its time, perhaps our EDEX system met the same fate.